\documentclass[superscriptaddress,secnumarabic ,
amssymb,amsmath,nobibnotes,twocolumn,showkeys, showpacs,aps,prd,nofootinbib]{revtex4}%
\usepackage{graphicx}
\usepackage{epsf}
\usepackage{bm}
\usepackage{amsmath}
\usepackage{amsfonts}
\usepackage{amssymb}
\usepackage{epstopdf}
\usepackage{natbib}
\usepackage{color}%
\providecommand{\U}[1]{\protect\rule{.1in}{.1in}}
\newcommand{\be}{\begin{equation}}
\newcommand{\ee}{\end{equation}}

\newcommand{\mincir}{\raise
-3.truept\hbox{\rlap{\hbox{$\sim$}}\raise4.truept\hbox{$<$}\ }}
\newcommand{\magcir}{\raise
-3.truept\hbox{\rlap{\hbox{$\sim$}}\raise4.truept\hbox{$>$}\ }}

\begin{document}

\title{Cosmological solutions and finite time singularities in  Finslerian geometry}
\author{Nupur Paul}
\email{nnupurpaul@gmail.com}
\affiliation{Department of Mathematics, Jadavpur University, Kolkata $-$ 700032, West Bengal, India}

\author{S. S. De}
\email{ssddadai08@rediffmail.com}
\affiliation{Department of Applied Mathematics, University of Calcutta, Kolkata $-$ 700009, India}


\author{Farook Rahaman}
\email{rahaman@associates.iucaa.in}
\affiliation{Department of Mathematics, Jadavpur University, Kolkata $-$ 700032, West Bengal, India}
\begin{abstract}

We consider a very general scenario of our universe where its geometry is characterized by the Finslerian structure on the underlying spacetime manifold, a generalization of the Riemannian geometry. Now considering a general energy-momentum tensor for matter sector, we derive the gravitational field equations in such spacetime. Further, to depict the cosmological dynamics in such spacetime  proposing an interesting equation of state identified by a sole parameter $\gamma$ which for isotropic limit is simply the barotropic equation of state $p= (\gamma- 1) \rho$ ($\gamma \in \mathbb{R}$ being the barotropic index), we solve the background dynamics. The dynamics offers several possibilities depending on this sole parameter as follows $-$ (i) only an exponential expansion, or (ii) a finite time past singularity (big bang) with late accelerating phase, or (iii) a nonsingular universe exhibiting an accelerating scenario at late time which finally predicts a big rip type singularity. We also discuss several energy conditions and the possibility of cosmic bounce. Finally, we establish the first law of thermodynamics in such spacetime.

\end{abstract}
\maketitle

 \section{Introduction and Preliminaries}
 \label{intro}
To understand the current accelerating expansion history of the universe, mainly two different approaches are
considered. The first one is to introduce some dark energy fluid in the framework of general relativity, while the later
introduces some modifications into the gravitational sector. Now, considering the dark energy concept, observations
favor the existence of $\Lambda$CDM cosmology where $\Lambda$ is the cosmological constant and it is considered to be responsible for
this accelerated expansion. However, this cosmology has one serious issue known as cosmological constant problem
which inspired several authors to consider some time varying dark energy fluid in the name of quintessence, phantom,
Chaplygin gas and so on, although it should me mentioned that they introduced some other issues, for instance the
cosmic coincidence problem and so on. Probably, both the cosmological constant as well as the time varying dark
energy models inspired cosmologists to introduce the modifications into the gravitational sector which finally appeared
with a large number of gravitational modifications, such as $f (R)$, $f (T )$, etc. However, the basic problem in all such
modified gravity theories is to first fix a functional form of $f (R)$ and $f (T )$ and then we study their viability with
the theoretical bounds and finally with the observational data\footnote{Exactly what we do while dealing with several dark energy fluids.}. That means essentially in these two formalisms, we
start with some phenomenological grounds. Thus, no one can exclude the possibility of a new dark energy fluid or some
new modified gravity theory, or even some new direction of research which may positively account of the current
astrophysical and cosmological issues.

In this article we have discussed the evolution of the universe in the context of Finslerian geometry which is a
generalization of the traditional Riemannian geometry \cite{chern, bao}\footnote{We remark that this geometry was in fact suggested by Riemann himself in his book [20]} . The basic concept in such theories is to consider the
violation of Lorentz symmetry which is one of the common features in the quantum gravitational regime and it may
introduce some new directions in the modern cosmological research. The metric in this space is defined by a norm
$F (x, y)$ (where $y \in T_x M$ is the tangent vector on a spacetime point $x$) on a tangent bundle of the spacetime instead
of an usual inner product structure on the corresponding spacetime, for a detailed description we refer \cite{PCS, Vacaru}. It
has been found that during last couple of years a considerable attention has been paid on this extended geometry
to address some issues related to astrophysics and cosmology \cite{fn1, fn2, fn3, fn4, fn5, fn6, Li2014, XWC, FR1, FR2}, specifically, it has been argued that the dark
matter and the dark energy problem can be addressed in such a context \cite{fn2,fn3,fn6}. So, essentially, the cosmology in
this space-time will be worth exploring for new physical results.

A Finslerian structure on a smooth $4-$dimensional manifold $M$ is defined on the tangent bundle $TM$ of  $M$ by Finsler metric $F=\tilde {T}M = TM - \{0\}$ and $F$ is smooth on $\tilde TM$. In addition $F$ is positively homogeneous of degree one with respect to $(y^a)$, i.e. $F(x,ky) = kF(x,y)$, $k > 0$. The Finslerian metric tensor $g_{ij}$ is given by
\be
g_{ij} = \dfrac{1}{2}\dfrac{\partial^2F^2(x,y)}{\partial y^i \partial y^j}
\ee
defined on $\tilde TM$.
The causality in Finsler spacetime can be given by the metric function $F$. A positive, zero or negative value of $F$ corresponds to timelike, null or space-like curves. In many cases it is useful that a Lorentz signature to be taken under consideration.
A non-linear connection $N$ on $TM$ is a distribution on $TM$, supplementary to the vertical distribution $V$ on $TM$ as
\be
T_{(x,y)}(TM) = N_{(x,y)}\oplus V_{(x,y)}.
\ee
A non-linear connection can be defined as
\be
N^a_j = \dfrac{\partial G^a}{\partial y^j},
\ee
where $G^a$ is given by
\be
G^a = \frac{1}{4} g^{aj}\left(\frac{\partial^2F^2}{\partial y^j \partial y^k}y^k - \partial_jF^2 \right).
\ee
The geodesic equation for the Finsler space follows from the Euler-Lagrange equations
\be
\frac{d}{ds}\left( \frac{\partial F}{\partial y^a}\right) - \frac{\partial F}{\partial x^a} = 0,
\ee
and it has the form
\be
\frac{dy^a}{ds} + 2G^a(x,y) = 0.
\ee

A  complete description of Finslerian structure of spacetime manifold $M$ is usually given with a metric function $F$, a nonlinear connection $N$ and compatible (metrical) connections in the framework of a tangent bundle $TM$ of spacetime \cite{Vacaru}. However in some cases we can restrict our consideration in order to describe the local anisotropic ansantz of gravitational field equations on a base $4-$dimensional manifold \cite{fn1}. In such a case we are able to compare different scenarios of Riemannian-type cosmological models as FLRW with Finslerian ones, e.g \cite{fn7}.

Thus, being motivated by the new generalized spacetime governed by a Finslerian metric, in the present work we
have tried to address the current evolution of the universe. Considering the background matter distribution to be
anisotropic we wrote the gravitational equations in such spacetime. Then we  introduce an interesting equation of state which for the isotropic limit simply assumes the barotropic equation of state. We show that under such equation of state, the field equations can be analytically solved which depending on the barotropic index of the fluid exhibits
several cosmological issues. We found that depending on the barotropic index of the fluid the cosmological solution
can predict a finite time past singularity and currently it can exhibit an accelerating scenario of our universe. On the
other hand, the model can offer a singularity free universe in the past while it can not escape from a finite time future singularity. Moreover, we also show that the violation of the energy conditions can lead to the cosmic bounce in such geometric structure. Finally, we establish that the first law of thermodynamics can hold under a simple energy conservation relation. Overall, the current picture offers some interesting possibilities.

We have organized the paper as follows: In section \ref{field-equations-Finsler}, we have presented the modified field equations in  Finsler space-time. Section \ref{the_model} deals with a toy model of an equation of state connecting the anisotropic matter distribution and evaluates the cosmological solutions. In \ref{energy}, we briefly describe the energy conservation relation and the first law of thermodynamics. Finally, we close our work in section \ref{discussion} with a brief summary.

\section{Gravitational equations in Finsler space-time}
\label{field-equations-Finsler}

To explain the dynamics of the universe, one desires to introduce a metric specifying
the geometry of the space-time which connects with its matter distribution by the Einstein's gravitational equation.
In order to realize the dynamics, we consider that our universe is described
by the Finsler metric which is of the form \cite{Li2014}

\begin{equation}\label{metric}
F^2=y^ty^t-R^2(t)y^ry^r-r^2R^2(t)\bar{F}^2(\theta,\phi,y^\theta, y^\phi),
\end{equation}
which has been taken inspired by the  known fact that at large scales our universe is
well described by the flat Friedmann-Lema\^{i}tre-Robertson-Walker (FLRW) line element.
We construct a cosmological model by inserting  $\bar{F}^2$  as a  quadric in $ y^\theta$ and $y^\phi $.
Note that   the   present Finsler space (\,for the case $\bar{F}^2$ as quadric in $ y^\theta$ and $y^\phi $\,)
can be obtained  from a Riemannian manifold  $( M, g_{\mu \nu}(x))$ as we have
   \[F(x,y) =\sqrt{g_{\mu \nu}(x)y^\mu y^\nu }.\]

One can notice  that, this is a semi-definite Finsler space-time and consequently, one can use the covariant
derivative of the Riemannian space. It is to be noted that the  Bianchi identities
overlap with those of the Riemannian space (being the
covariant conservation of Einstein tensor). Since the present Finsler space
reduces to the Riemannian space, therefore,   the gravitational
field equations can be obtained readily.
The  base manifold of the Finsler space regulates the gravitational field
equation in Finsler space and the fiber coordinates $y^i$ play the role of
the velocities of the cosmic components i.e. velocities in the energy
momentum tensor. Hence, one can derive
 the gravitational field equations in Finsler space  from the Einstein gravitational field equation in
 the Riemannian space-time with the metric (\ref{metric}) in which the metric $\bar{g}_{ij}$ is given by (see Appendix A)

\begin{align}
\bar{g}_{ij} = \mbox{diag}(1,\sin^2 \sqrt{\lambda}\,\, \theta),\nonumber
\end{align}
i.e.

\begin{align}
g_{\mu\nu }= \mbox{diag}(1,-R^2(t),-r^2R^2(t),-r^2R^2(t) \sin^2 \sqrt{\lambda}\,\, \theta),\nonumber
\end{align}
where $\lambda > 0$  and it plays a very crucial role in the derived field equations in Finsler spacetime and hence in the background cosmology.
The two dimensional Finsler space $\bar{F}$ is specified as a
constant flag curvature space, that is it is assume that
$\bar{R}\mbox{ic}=\lambda$. This Ricci scalar ``$\bar{R}$ic'' is that of the
Finsler structure $\bar{F}$ and two dimensional Finslerian
structure is specified by the constancy of the flag curvature.
This flag curvature of Finsler space is in fact, the generalization of
the sectional curvature of Riemannian space. It will be apparent latter
that the solution of vacuum field equation must lead to the constancy
of the flag curvature with its value $\lambda = 1$. But for more general
case of Finsler  structure $\bar{F}$, it can be specified by the constant
flag curvature having any real value for $\lambda $. In fact, for
$\lambda = 1$, we can get the usual Friedmann equations of the FLRW universe.

Let us now assume the general energy-momentum tensor for the matter sector as

\begin{equation}\label{em-tensor}
T^\mu_\nu =(\rho +p_t)u^\mu u_\nu -p_tg^\mu_\nu+(p_r-p_t)\eta^\mu \eta_\nu,
\end{equation}
where $u^\mu u_\mu = -\eta^\mu \eta_\mu = 1$, $p_r$, $p_t$ are respectively
denote the pressures of the anisotropic fluid in the radial and transversal directions.
The modified gravitational  field equations  in Finsler space-time are obtained as  (see Appendix B)

\begin{eqnarray}
8\pi_F G\rho &=& \frac{3\dot{R}^2}{R^2}+\frac{\lambda-1}{r^2R^2},\label{F1}\\
8\pi_F G p_r &=& -\frac{2\ddot{R}}{R}-\frac{\dot{R}^2}{R^2}-\frac{\lambda -1}{r^2R^2},\label{F2}\\
8\pi_F G p_t &=& -\frac{2\ddot{R}}{R}-\frac{\dot{R}^2}{R^2}.\label{F3}
\end{eqnarray}
Note that,  $\lambda = 1$ implies $p_r=p_t$, that means, it helps to recover the gravitational
field equations for flat FLRW universe. On the
other hand, if we put $p_r= p_t$ in the above field equations, we readily find $\lambda= 1$. Hence,
we find that, $\lambda= 1 \Longleftrightarrow$ flat FLRW universe. Moreover, one can easily see
that when $t\longrightarrow \infty$, we again find that both the anisotropic pressure
components become equal and the usual Friedmann equations in the spatially flat FLRW
universe in presence of  a perfect fluid with energy-momentum
tensor $T_{\mu \nu}= (p+\rho) u_{\mu} u_{\nu}+ p g_{\mu \nu}$ are recovered.

Now, introducing the Hubble parameter $H= \dot{R}/ R$,
the field equation  (\ref{F1}) can be written as
(for simplicity, from now we work in the units where $8 \pi_F G= 1$)

\begin{eqnarray}\label{F1a}
\rho &=& 3 \left(H^2+ \frac{\lambda-1}{3\, r^2\, R^2}\right)\Longleftrightarrow~ \Omega+ \Omega_k = 1,
\end{eqnarray}
where $\Omega= \rho/3H^2$ is the density parameter representing the matter sector, and in
compared to the Friedmann universe, the quantity $\Omega_k= -\frac{(\lambda-1)}{3 (r\,R\,H)^2}$
can be looked as the density parameter for the scalar curvature in the Finslerian geometry.
Also, for both the directions, the Raychaudhuri equation can be written as

\begin{eqnarray}
\dot{H} &=& -\frac{1}{2} (\rho+ p_r),\label{F2a}\\
\dot{H} &=& -\frac{1}{2} (\rho+ p_t)+ \frac{\lambda-1}{2\, r^2\,R^2},\label{F3a}
\end{eqnarray}
 where again we note that, for $\lambda= 1$,
 it reduces to only one equation.

\section{Evolution and Dynamics: A toy model}
\label{the_model}
In general, the equation of state of this anisotropic fluid takes a general form $f (p_r,~p_t,~\rho)= 0$. The exact form of the equation of state is not known and hence, it is a challenge for modern cosmology to derive the cosmological evolution correctly. Still, we adopt mainly two possible ways. One is to assume a very simple formulation of the equation of state in order to derive the evolutionary parameters so that we can match them with the observational data, and on the other hand, the reconstruction of any quantity from observed data is of worth exploring. However, in the present work,  we adopt the first possibility, and thus, we start with the following equation of state
\begin{equation}\label{eos}
\gamma (p_t- p_r)+p_r =(\gamma-1) \rho 
\end{equation}
where  $\gamma \in \mathbb{R}$ is simply a constant. The essence of this equation of state is that
for $p_t= p_r= p$ (say), the equation of state in (\ref{eos})
is simply reduced to $p= (\gamma - 1) \rho$, representing the barotropic equation of state. Further, we notice that, for $\gamma= 0$, equation (\ref{eos}) implies $p_r= -\rho$, and hence $p_t= - 3 \dot{R}^2/R^2$. We are interested in the cosmological solutions for the above choice of the equation of state.  Now, using the field equations (\ref{F1}), (\ref{F2}), (\ref{F3}), we can exactly solve the scale factor $R$ as

\begin{align}
R &= R_0 \left[1+ \frac{3\,\gamma}{2} H_0 (t- t_0)\right]^{\frac{2}{3\, \gamma}},~(\gamma \neq 0)\label{scale-factor}\\
R &= R_0 \exp\left(H_0 (t-t_0)\right),~~~~(\gamma = 0)\label{sf2}
\end{align}
where $t_0$, $H_0$ are respectively the present cosmic time and present day value of the Hubble parameter, and it is worth noting that the solutions obtained in Eqns (\ref{scale-factor}), (\ref{sf2}) exactly match with the solutions obtained for the isotropic matter distribution in the FLRW geometry with the equation of state $p= (\gamma -1 ) \rho$. Consequently, the Hubble parameter can be derived as

\begin{eqnarray}
H &=& \frac{H_0}{1+ \frac{3\, \gamma}{2} H_0 (t- t_0)},~~~~(\gamma \neq 0)\label{Hubble}\\
H &=& H_0= \mbox{Constant},~~~~~(\gamma = 0)\label{Hubble2}
\end{eqnarray}

Therefore, it is clear that for $\gamma= 0$ we realize an exponential expansion of
the universe. On the other hand, we can divide $\gamma \neq 0$  into the following two conditions
when $\gamma > 0$ and $\gamma < 0$.

\subsection{The case for $\gamma > 0$}

In this case, the solutions for the scale factor and the Hubble parameter take the forms as
in equations (\ref{scale-factor}) and (\ref{Hubble}). The solutions offer the following
scenario of our universe.

\begin{eqnarray}
\label{BB-singularity}
\mbox{At}~t_f = t_0- \frac{2}{3\,H_0\, \gamma},~~~~~~~~~~~\nonumber\\
~~\lim_{t \rightarrow t_f} R (t)= 0,~~\mbox{and}~~\lim_{t \rightarrow t_f} H (t)= \infty\, ,
\end{eqnarray}
which clearly shows that the universe attains a big bang singularity in the past ($t_f < t_0$).
On the other hand, at late time we find that

\begin{eqnarray}
\label{future-evolution}
\lim_{t \rightarrow \infty} R (t)&=& \infty,~~~~\mbox{and}~~~~\lim_{t \rightarrow \infty} H (t)= 0.
\end{eqnarray}

Now, introducing the deceleration parameter $q= -1-\dot{H}/H^2$,
we find that, for this cosmological solution one has $q= -1 + 3\gamma/2$, which
represents an accelerating universe (i.e. $q< 0$) for $\gamma < 2/3$. Thus,
the model with the equation of state in (\ref{eos}) presents a model of
our universe which predicts the big-bang singularity (a finite time
singularity, but independent of $\gamma$), and describes an accelerating
universe for $\gamma < 2/3$.

\subsection{The case for $\gamma < 0$}

Now, we consider the cosmological solutions for $\gamma< 0$. For a clear image let us consider
$\gamma= -\alpha$ (where $\alpha> 0$). We rewrite  the scale factor and the Hubble parameter as

\begin{eqnarray}
R&=& \frac{R_0}{\Bigl[1- \frac{3\,\alpha}{2} H_0 (t-t_0)\Bigr]^{\frac{2}{3\alpha}}},\label{SF-emergent}\\
H &=& \frac{H_0}{1- \frac{3\, \alpha}{2} H_0 (t- t_0)}.
\end{eqnarray}

We find that the scale factor can not be zero at any finite cosmic time in the past evolution of the universe, in other words it gives a solution to the nonsingular universe that has been proposed in several contexts in modern cosmology with great interests, see Refs. \cite{BV1988, EM2004, Ellis2004, Mulryne2005, Nunes2005,Lidsey2006,Banerjee2007,Banerjee2008},  but on the other hand, from the solution of the scale factor in eqn. (\ref{SF-emergent}), it is seen that it diverges
at some finite cosmic time in future, that means for $t_s= t_0+ \frac{2}{3\alpha H_0}$, $R(t) \longrightarrow \infty$.
Thus, we find that this cosmological solution gives a realization of a nonsingular universe in the past which consequently predicts a ``big rip'' singularity.

Now, we find that we arrive at two different cosmological scenarios,
one which starts with big bang but at late time it does not have any singularity, the other
has a nonsingular nature in past but at future it has a big rip singularity. Therefore, in the
next section we constrain the model parameters for the first cosmological solution (i.e. for $\gamma > 0$).

Let us look at a particular case when
both the pressure components satisfy barotropic equations of state, that means, $p_r= \omega_r \rho$, $p_t= \omega_t \rho$, and  where we assume that $\omega_r$ and $\omega_t$ are the constants. Now, it is easy to see that under such conditions, one can exactly solve the energy density using the above two equations, where the explicit form for $\rho$ is
\begin{equation}
\rho= \rho_0 \left( \frac{r^{2(\omega_t-\omega_r)}}{R^{(3+\omega_r+ 2\omega_t)}}\right)
\end{equation}
where $\rho_0 > 0$ is an integration constant and it recovers the standard evolution equations in Friedmann cosmology for $\omega_r = \omega_t$.
Additionally, the energy conditions in this spacetime can be written in the following way:\\

\textbf{WEC:} $\rho \geq 0$, and $\rho+ p_r \geq 0$, $\rho+ p_t \geq 0$\\

In the Finsler space-time, using the gravitational equations  (\ref{F1}), (\ref{F2}), (\ref{F3}), the conditions respectively reduced to  the following inequalities as \\

$$3\left(\frac{\dot{R}^2}{R^2}\right)+ \frac{\lambda-1}{r^2\,R^2} \geq 0,\,\,~~\frac{\ddot{R}}{R} \leq \frac{\dot{R}^2}{R^2},~~\mbox{and}~~\frac{\ddot{R}}{R} \leq \frac{\dot{R}^2}{R^2}+ \frac{\lambda-1}{2\,r^2\,R^2}$$

\textbf{SEC:} $\rho+ p_r \geq 0$, $\rho+ p_r+ 2 p_t \geq 0$\\

In this case, the conditions are reduced to $$\frac{\ddot{R}}{R} \leq \frac{\dot{R}^2}{R^2},~~~~\mbox{and}~~~~\frac{\ddot{R}}{R} \leq 0$$

\textbf{NEC:} $\rho+ p_r \geq 0$, and $\rho+ p_t \geq 0$\\

Here, using the gravitational equations, these conditions respectively reduced to

$$\frac{\ddot{R}}{R} \leq \frac{\dot{R}^2}{R^2},\,\,~~\mbox{and}~~\frac{\ddot{R}}{R} \leq \frac{\dot{R}^2}{R^2}+ \frac{\lambda-1}{2\,r^2\,R^2}$$

\textbf{DEC:} $\rho \geq 0$, and $-\rho \leq p_r \leq \rho$, $-\rho \leq p_t \leq \rho$\\

In a similar way, the inequalities respectively reduced to

$$3\left(\frac{\dot{R}^2}{R^2}\right)+ \frac{\lambda-1}{r^2\,R^2} \geq 0,\,\,$$

\begin{align}
-2\left(\frac{\dot{R}^2}{R^2}\right)- \frac{\lambda-1}{r^2\,R^2}\,\leq \, \frac{\ddot{R}}{R}\, \leq \, \frac{\dot{R}^2}{R^2},\,\,~~~~~~~~~\nonumber\\~~\mbox{and}~~-2\left(\frac{\dot{R}^2}{R^2}\right)- \frac{\lambda-1}{r^2\,R^2} \, \leq \, \frac{\ddot{R}}{R} \, \leq \, \frac{\dot{R}^2}{R^2}+ \frac{\lambda-1}{2\,r^2\,R^2}\nonumber
\end{align}
Clearly the term $\left(\frac{\lambda-1}{r^2 R^2}\right)$ makes a significant contribution to the energy conditions, where we note that for $\lambda= 1$, the above conditions are simply reduced to the energy conditions as we find for a flat FLRW universe with a perfect fluid distribution.

\subsection{Bouncing conditions}

In this section we will seek for bouncing conditions in this spacetime. Since the expansion scalar in this spacetime is defined to be $H = \dot{R}/R$, therefore one may recall the bounce condition  \cite{CDS}

\begin{eqnarray}
H (t_b) = 0,\,\,\,\,\mbox{and}\,\,\,\,\dot{H}(t_b) > 0 \label{bounce_conditions}
\end{eqnarray}
where $t_b$ is the time where universe has bounced. Actually, the above conditions can also be written as $\dot{R} (t_b) = 0$ and $\ddot{R} (t_b)> 0$. From the above energy conditions, following the bounce conditions, an immediate solution for the bouncing universe is that $\lambda > 1$.

By using bounce conditions \eqref{bounce_conditions} in (9), (10), (11) we get for the energy conditions

\begin{itemize}
 \item WEC : $\rho+p(r)<0, \rho+p(t)<0, \lambda>1 $, so WEC is violated.

 \item SEC :  $\rho+p(r)<0, \rho+p(r)+2p(t)<0$, so SEC is violated.

 \item NEC and DEC are also violated.
\end{itemize}
These cases are necessary conditions in order to have a cosmic bounce.


%
%
%
%
%
%
%

\section{Energy Conservation relations}
\label{energy}
In this section we shall devote our discussions on the energy conservation relation. Let us
propose the energy conservation equation as follows

\begin{equation}\label{conservation}
d(\rho V) = -P dV -V F_a dr.
\end{equation}
The additional term in R.H.S. is due to the anisotropic force $F_a$ given by
\begin{equation}
F_a = \frac{2(p_t-p_r)}{r} =\frac{2(\lambda -1)}{8 \pi_FG}\frac{1}{r^3R^2},
\end{equation}
and the pressure $P$ is the average pressure which is given by
\begin{equation}\label{pressure}
P=\frac{p_r+p_t+p_t}{3}=\frac{p_r}{3}+\frac{2p_t}{3}.
\end{equation}
The proposed conservation equation   (\ref{conservation}) can be written as

\[Vd\rho +\rho dV+PdV+VF_adr=0,
\]
or, alternatively as
\[d\rho+\rho \frac{dV}{V}+P\frac{dV}{V}+F_adr=0,
\]
which then turns into

\[\frac{\partial \rho}{\partial t}dt+\frac{\partial \rho}{\partial
r}dr+(\rho+ P)\frac{dV}{V}+F_a dr = 0,\]
which again can be recast as

\begin{align}
dt\left(\frac{\partial \rho}{\partial t}+(\rho
+P)\frac{\dot{V}}{V}\right)+dr\left(\frac{\partial \rho}{\partial
r}+F_a\right)=0,\nonumber
\end{align}
which clearly reports
the following two separate equations:
\begin{eqnarray}
\frac{\partial \rho}{\partial t} + 3H(\rho+P) &= & 0,\label{cons1}\\
\frac{\partial \rho}{\partial r}& = & -F_a.\label{cons2}
\end{eqnarray}
Note that, the first equation is the usual energy conservation equation for the homogeneous and isotropic universe with the effective pressure $P$. The second equation can be derived from the above field equations (\ref{F1}), (\ref{F2}), (\ref{F3}). That means, the proposed conservation relation (\ref{conservation}) is consistent and well motivated with the energy conservation relation in the context of general relativity.

Also, if we define a pressure $P$ as the weighted average of $p_t$ and $p_r$, i.e. if $P= \gamma p_t-(\gamma- 1) p_r$ (note that the sum of the weight is $\gamma- (\gamma-1)=1$) then we see that the present equation of state is the barotropic equation of state $P=(\gamma- 1)\rho$. In equation (\ref{pressure}) we have taken $P$ as the simple arithmetic average of the radial and transverse pressure, i.e. the pressure in three orthogonal directions.
 This case corresponds $\gamma =\frac{2}{3} $.
With this, the conservation equation becomes
\begin{equation}\label{cons-sp}
d(\rho V)+PdV+\left(\gamma-\frac{2}{3}\right)(p_r-p_t)dV+\tilde{F}dr=0
\end{equation}
where, $\tilde{F}=VF_a $ is an anisotropic force. In fact, the equation (\ref{cons-sp}) can be written as \begin{equation}
dU+dW=dQ=0,
\end{equation}
where $U=\rho V$ is the internal energy, $W$ is the work done and $Q$ is the heat transfer and using these terminology, eqn. (\ref{cons-sp}) is nothing but the following equation
\begin{equation}
dW= PdV +\left(\gamma- \frac{2}{3}\right)(p_r-p_t)dV+\tilde{F}dr.
\end{equation}
The first two terms in the right hand side being the change of  work from pressures and the third being that due to the anisotropic force. This equation represents the first law of thermodynamics for the case of adiabatic heat transfer.

\section{Concluding remarks}
\label{discussion}

In this work we considered the spacetime of our universe described by the Finsler geometry, a generalization of the Riemannian geometry \cite{chern, bao}. A number of studies  \cite{fn1, fn2, fn3, fn4, fn5, fn6, Li2014, XWC, FR1, FR2} in this framework have been performed in order to offer some explanations on some recent astrophysical and cosmological evidences. Now, considering a general matter distribution which by construction stands for an anisotropic matter distribution, we rewrote the modified gravitational field equations. We solved the dynamics of the universe for a simple equation of state of the matter sector (see eq. (\ref{eos})) characterized by a sole parameter $\gamma$ and  which is motivated from the fact that in case of a perfect fluid matter distribution the barotropic equation of state, is recovered. Since $\gamma$ has been kept free for our analysis, so depending on it, we found three distinct cosmological scenarios with current interests:

\begin{itemize}
	\item For $\gamma = 0$, we realize an exponential expansion of the universe. But, we do not have any such other information for such specific value of $\gamma$.
	
	\item $\gamma > 0$ is of special importance in this study since the model perfectly indicates big bang singularity in the past and finally there is an accelerating phase for $\gamma < 2/3$. So, this model can trace back the early phase of the universe as well as stands for the late time accelerated phase.
	
	\item The case $\gamma < 0$ is also interesting because the model in this case does not encounter any finite time singularity in the past. That means, an existence of nonsingular universe is supported as observed in other cosmological theories \cite{BV1988, EM2004, Ellis2004, Mulryne2005, Nunes2005,Lidsey2006,Banerjee2007,Banerjee2008}. Additionally, the late accelerating universe is also realized, but finally at some finite future time the scale factor diverges (Big rip singularity). Therefore, the model starts from a nonsigular phase, exhibits an accelerating universe and finally  hints  for the divergence in the scale factor at finite time in future.
	
\end{itemize}

Further, we show that the violation of the energy conditions can lead to the cosmic bounce. We also further established the energy conservation relation which show that the model is in well agreement with the first law of thermodynamics.

Thus, in summary considering the background spacetime of our universe as Finsler geometry instead of Riemannian geometry it is observed  that a simple barotropic like fluid can offer some interesting cosmological solutions which accommodate the past and present scenarios of the universe evolution. Thus, as a beginning, the current study may be considered as a viable one for the next complicated models in Finsler cosmology.

\section*{Acknowledgments}

FR would like to thank the authorities of the Inter-University Centre
for Astronomy and Astrophysics, Pune, India for providing research facilities.   FR is also grateful to DST-SERB and DST-PURSE,  Govt. of India for financial support.
We wish  to thank  Panayiotis C. Stavrinos and Supriya Pan for helpful discussion.

\appendix

\section{Basics of Finslerian geometry}
\label{appA}

We construct a cosmological model by inserting  $\bar{F}^2$ in the following form

\begin{equation}\label{metric1a}
\bar{F}^2=y^\theta y^\theta+f(\theta,\phi)y^\phi y^\phi.
\end{equation}
Here,  \[ \bar{g}_{ij}=\mbox{diag}(1,f(\theta,\phi)),~~\mbox{and}~~\bar{g}^{ij}=\mbox{diag}(1,\frac{1}{f(\theta,\phi)});~~~[i,j=   \theta  ,
\phi ]\]

The geodesic equations in Finsler space-time are

\[\frac{d^2x^\mu}{d\tau^2} + 2G^\mu =0,\]
where $G^\mu = \frac{1}{4} g^{\mu \nu}
\left( \frac{\partial^2 F^2}{\partial x^\lambda \partial y^\nu} y^\lambda - \frac{\partial F^2}{\partial x^\nu} \right)$
are the geodesic spray coefficients which calculated from $\bar{F}^2$ as

\[ G^\theta=-\frac{1}{4}\frac{\partial f}{\partial \theta} y^\phi y^\phi\]

\[ G^\phi=\frac{1}{4f}\left(2\frac{\partial f}{\partial
\theta}y^\phi y^\theta+\frac{\partial f}{\partial
\phi}y^\phi y^\phi\right)\]
Hence, one can find

\begin{eqnarray}\label{barF}
&&\bar{F}^2\bar{R}\mbox{ic} =y^\phi y^\phi \Bigl[-\frac{1}{2}\frac{\partial^2
f}{\partial \theta^2}  +\frac{1}{2f}\frac{\partial^2 f}{\partial
\phi^2}-\frac{1}{2}\frac{\partial}{\partial
\phi}\left(\frac{1}{f} \frac{\partial f}{\partial
\phi}\right)-\nonumber\\
&&\frac{1}{4f} \left(\frac{\partial f}{\partial
\theta}\right)^2 \Bigr] -y^\phi y^\phi \Bigl[ \frac{1}{2f^2} \left(\frac{\partial f}{\partial
\phi}\right)^2+\frac{1}{4f}\frac{\partial f}{\partial
\phi}\frac{1}{f}\frac{\partial f}{\partial \phi}\nonumber\\
&&+\frac{\partial
f}{\partial \theta}\frac{1}{2f}\frac{\partial f}{\partial
\theta}-\frac{1}{4f^2}\left(\frac{\partial f}{\partial
\phi}\right)^2\Bigr]\nonumber\\
&&+y^\theta
y^\theta\left[-\frac{1}{2}\frac{\partial}{\partial
\theta}\left(\frac{1}{f} \frac{\partial f}{\partial
\theta}\right)-\frac{1}{4f^2}\left(\frac{\partial f}{\partial
\theta}\right)^2\right]\nonumber\\
&&+ y^\phi y^\theta\left[\frac{1}{f}\frac{\partial^2f}{\partial \theta \partial
\phi} - \frac{1}{f^2} \left(\frac{\partial f}{\partial
\theta}\right)  \left(\frac{\partial f}{\partial
\phi}\right) \right]\nonumber\\
&&-y^\phi y^\theta \left[  \frac{1}{2}\frac{\partial}{\partial
\theta}\left(\frac{1}{f} \frac{\partial f}{\partial
\phi}\right)+\frac{1}{2}\frac{\partial}{\partial
\phi}\left(\frac{1}{f} \frac{\partial f}{\partial
\theta}\right)\right]
\end{eqnarray}

The coefficient of  $y^\phi y^\theta  $ is zero iff, $ f$  is independent of $\phi$  i.e.
  \begin{equation}
  f(\theta, \phi) =f(\theta)
  \end{equation}
Note that other  coefficients of $y^\theta y^\theta $ $\&$  $y^\phi y^\phi $ are non zero.
\\

For the above consideration, Eq.(\ref{barF}) yields
\[ \bar{F}^2\bar{R}\mbox{ic}=\left[-\frac{1}{2f}\frac{\partial^2
f}{\partial \theta^2}  +\frac{1}{4f^2}\left(\frac{\partial f}{\partial
\theta}\right)^2\right] (y^\theta y^\theta+f y^\phi y^\phi)\]
This yields  $\bar{R}\mbox{ic}$ as
\begin{equation}
\bar{R}\mbox{ic}= -\frac{1}{2f}\frac{\partial^2
f}{\partial \theta^2}  +\frac{1}{4f^2}\left(\frac{\partial f}{\partial
\theta}\right)^2.
\end{equation}
It is obvious that $\bar{R}\mbox{ic}$ may be a constant,  say $\lambda$, or a function of $\theta$.
\\

For $\bar{R}\mbox{ic} = \lambda$, one can find  Finsler structure $\bar{F}^2$ as

$$\bar{F}^2= \left\{\begin{array}{ll} y^\theta y^\theta  + A \sin^2(\sqrt{\lambda} \theta )y^\phi y^\phi, & \mbox{if}~~\lambda > 0\\
 y^\theta y^\theta  + A \theta^2 y^\phi y^\phi, & \mbox{if}~~\lambda = 0,\\ y^\theta y^\theta  + A \sinh^2(\sqrt{-\lambda} \theta )y^\phi y^\phi, & \mbox{if}~~ \lambda < 0.
 \end{array}\right.$$
We note that one can take  $A$  as unity  without any loss of generality.
Thus, we get the following form of the  Finsler structure (\ref{metric})  as

\begin{eqnarray}
&&{F}^2  =  y^ty^t - R^2(t) y^ry^r -r^2R^2(t) y^\theta y^\theta \nonumber\\
&& - r^2R^2(t)\sin^2 \theta y^\phi y^\phi + r^2R^2(t)\sin^2 \theta y^\phi y^\phi \nonumber\\
&&-r^2R^2(t)\sin^2(\sqrt{\lambda} \theta )y^\phi y^\phi,\nonumber\\
&&= \alpha^2 + r^2 R^2(t)(\sin^2 \theta - \sin^2(\sqrt{\lambda} \theta ))y^\phi y^\phi,\nonumber
\end{eqnarray}
which implies
\begin{equation}
{F}^2 = \alpha^2 + r^2 R^2(t)\chi (\theta)y^\phi y^\phi
\end{equation}
where  $\chi (\theta) = \sin^2 \theta - \sin^2(\sqrt{\lambda} \theta )$, and $\alpha$ is a Riemannian metric.
Therefore, choosing  $b_\phi= r R(t)\sqrt{ \chi (\theta)} $, we get

\begin{equation}
F = \alpha \phi(s)~,~\phi(s) = \sqrt{1+s^2}
\end{equation}
where $s =\frac{(b_\phi y^\phi)}{\alpha} = \frac{\beta}{\alpha}$, and
$b_\mu = ( 0,0,0,b_\phi), b_\phi y^\phi = b_\mu y^\mu = \beta$, ($\beta$ is one form).
This readily shows that   $F$ is the metric of $(\alpha, \beta)$-Finsler space.
One can write  the killing equation $K_V(F) =0 $ in  Finsler space by using isometric transformations of Finsler structure   as \cite{Li2014}

   \begin{equation}
   \left(\phi(s)-s\frac{\partial \phi(s)}{\partial s}\right)K_V(\alpha) + \frac{\partial \phi(s)}{\partial s }K_V\beta) =0,
   \end{equation}
   where
\[
   K_V(\alpha) = \frac{1}{2 \alpha}\left(V_{\mu \mid \nu} +V_{\nu \mid \mu} \right)y^\mu y^\nu; \]\[
    K_V(\beta) =  \left(V^\mu \frac{\partial b_\nu}{\partial x^\mu } +b_{ \mu}\frac{\partial V^\mu}{\partial x^\nu} \right)y^\nu.\]
 The symbol  ``$\mid$'' means  the covariant derivative with respect to the Riemannian metric $\alpha$. Now,  we have
   \[K_V(\alpha)+ sK_V(\beta)=0, ~~\mbox{or}~~  \alpha K_V(\alpha)+  \beta K_V(\beta)=0,\]
  which gives
 \begin{equation}
 K_V(\alpha)= 0,~~\mbox{and}~~K_V(\beta)=0
 \end{equation}
 or
\begin{equation}
  V_{\mu \mid \nu} +V_{\nu \mid \mu} =0
  \end{equation}
and
\begin{equation}
V^\mu \frac{\partial b_\nu}{\partial x^\mu } +b_{ \mu}\frac{\partial V^\mu}{\partial x^\nu} =0.
\end{equation}

 Here, the second Killing equation constrains the first one (Killing equation of the Riemannian space).
     This takes responsibility  for breaking the symmetry (isometric) of the Riemannian space.

\section{Gravitational field equations}

The geodesic spray coefficients for the Finsler structure (\ref{metric}) are given by

\begin{equation}
G^t=\frac{1}{2} R \dot{R}\Bigl[r^2\bar{F}^2+y^ry^r  \Bigr]
\end{equation}
\begin{equation}\label{sp1a}
G^r= \left(\frac{\dot{R}}{R}\right)y^r y^t-\frac{r}{2}\bar{F}^2
\end{equation}
\begin{equation}\label{sp1b}
G^\theta=\bar{G}^\theta+y^\theta\left[\frac{y^r}{r}+\frac{\dot{R}}{R}\,y^t\right]
\end{equation}
\begin{equation}\label{sp1c}
G^\phi=\bar{G}^\phi+y^\phi\left[\frac{y^r}{r}+\frac{\dot{R}}{R}\,y^t\right]
\end{equation}
The Ricci scalar
\begin{eqnarray}\label{sp1d}
&&\mbox{Ric} \equiv  R_\mu^\mu = \frac{1}{F^2} \Bigg[ 2 \frac{\partial G^\mu}{\partial x^\mu} - y^\lambda \frac{ \partial^2 G^\mu}{\partial x^\lambda y^\mu} \nonumber\\
&& +
2\frac{G^\lambda \partial^2 G^\mu}{\partial y^\lambda y^\mu} -   \frac{\partial G^\mu}{\partial y^\lambda} \frac{\partial G^\lambda}{\partial y^\mu}\Bigg]
\end{eqnarray}
can be computed as

\begin{eqnarray}\label{sp1e}
&&F^2R\mbox{ic} =\bar{F}^2\Bigg[\bar{R}\mbox{ic}-1+ r^2\left( 2 \dot{R}^2 + R \ddot{R}\right)\Bigg]+ \nonumber\\
&&y^ry^r\left( 2 \dot{R}^2 + R \ddot{R}\right)-3 y^ty^t\left( \frac{\dot{R}^2}{R^2} + \frac{\ddot{R}}{R} - \dot{R}^2  \right)
\end{eqnarray}

For Ricci tensor $R\mbox{ic}_{\mu \nu} = \frac{  \partial^2 (\frac{1}{2} F^2Ric)}{\partial y^\lambda y^\mu}$, one can write the scalar curvature in Finsler geometry as $S= g^{\mu \nu} Ric_{\mu \nu}$, and explicitly in the following way


\begin{eqnarray}
S = - 3 \left[\frac{\ddot{R}}{R} + \frac{\dot{R}^2}{R^2} \right] - \frac{2(\lambda-1)}{r^2 R^2}
\end{eqnarray}	
An immediate observation shows that if we set $\lambda= 1$, then we find that $S$ is half of the Ricci scalar curvature in FRW universe. The modified Einstein field equations in Finsler space-time ($G_{\mu \nu} \equiv R\mbox{ic}_{\mu \nu} - \frac{1}{2} g_{\mu \nu} S = 8\pi_F G T_{\mu \nu}$) yield

\begin{equation}
G^t_t= 3 \left(\frac{\dot{R}^2}{R^2} \right)+\frac{\lambda-1}{r^2 R^2}=8\pi_F G T^t_t
\end{equation}
\begin{equation}
G^r_r=2\; \frac{\ddot{R}}{R}+ \frac{\dot{R}^2}{R^2}+\frac{\lambda-1}{r^2 R^2}=-8\pi_F G T^r_r
\end{equation}
\begin{equation}
G^\theta_\theta =G^\phi_\phi =2\; \frac{\ddot{R}}{R}+ \frac{\dot{R}^2}{R^2} = -8\pi_F GT^\theta _\theta = -8\pi_F GT^\phi_\phi
\end{equation}

As in Akbar-Zadeh \cite{ak}, with the modified Einstein tensor
($G_{\mu\nu} = Ric_{\mu\nu}-\frac{1}{2}g_{\mu\nu} S)$,\begin{equation}
\left(G_{\mu\nu}-8\pi G T_{\mu\nu}\right)|_M=0 \end{equation}
where $'|_M'$ represents this gravitational field equation restricted to the base space manifold $M$ of the Finslerian length element $F$. Here fibre  coordinates are supposed to be velocities of the cosmic components. Regarding the validity of gravitational field equations \cite{Li2014} it has been argued that these equations can be derivable from the gravitational dynamics for Finsler spacetime based on an action integral approach on the unit tangent bundle \cite{pf}. Of course, this derivation of the equation (\ref{sp1d}) from the gravitational field equation in Finsler spacetime as given by Pfeifer et. al  is only approximate, where the curvature tensor not belonging to the base space of the tangent bundle has been neglected. But the strong point is that the equation (\ref{sp1d}) is constructed by the geometrical invariant (Ricci tensor) which is insensitive to the connections. The equations for the present case are equations
 (\ref{sp1a}), (\ref{sp1b}) and (\ref{sp1c}).
These equations are the same as those which can be derived from the equivalent Riemannian spacetime with the with the metric tensor (\ref{metric}). Thus we see that although the equation (\ref{sp1d}) is ``approximately'' valid as it is restricted to the base space of the Finslerian length element F, this equation seems to be exact for the present case of FLRW spacetime with Finslerian perturbation $\beta=b_\mu b^\mu$.

\end{document}